# Development of high power quantum well lasers at RRCAT

T. K. Sharma[1,*], Tapas Ganguli[1], V. K. Dixit[1], S. D. Singh[1], S. Pal[1], S. Porwal[1], Ravi Kumar[1], Alexander Khakha[1], R. Jangir[1], V. Kheraj[2], P. Rawat[3], and A. K. Nath[1]

[1]Semiconductor Laser Section, Raja Ramanna Centre for Advanced Technology, Indore-452013(M.P)
[2]Applied Physics Department, M. S. University of Baroda, Vadodara-390001(Guj)
[3]School of Physics, Vigyan Bhawan, DAVV, Indore-452017(M.P)
[*]Email: tarun@cat.ernet.in

**Abstract**

We at RRCAT have recently developed high power laser diodes in the wavelength range of 740 to 1000 nm. A typical semiconductor laser structure is consisted of about 10 epilayers with different composition, thickness and doping values. For example, a laser diode operating at 0.8 μm has either GaAs or GaAsP quantum well as an active layer. The quantum well is sandwiched between AlGaAs wider bandgap waveguide and cladding layers. The complete laser structure is grown by metal organic vapour phase epitaxy technique and devices are fabricated through standard procedure using photolithography. We recently achieved about 5.3 Watt peak power at 853 nm. These laser diodes were tested under pulsed operation at room temperature for 500 nanosecond pulse duration with a duty cycle of 1:1000. Laser diode arrays consisting of 6-10 elements were also developed and tested for operation in pulsed mode at room temperature.

## 1. Introduction

Semiconductor lasers are the key components in opto-electronics and photonic integrated circuits and play an essential role in many swiftly expanding technologically important areas [1-3]. These are used in many applications like long-distance telephone conversations, wide-scale usage of compact disc technology in the interior of CD/DVD players, laser printing, optical communication which include overland, under sea long-haul systems of large capacity, intra-building, intra-ship and intra-aeroplane systems etc. Specifically, high power quantum well (QW) lasers have found a place in a wide variety of applications such as material processing, medical diagnosis, surgery, spectroscopy, and diode pumped solid state lasers including fiber lasers and amplifiers. Solid State laser pumping applications have long been one of the biggest motivations for the development of high power QW lasers. There is a large overlap between the available high power QW laser wavelengths and necessary pump wavelengths for many solid state lasers. Solid state lasers have certain attributes which are hard to achieve with diode lasers alone. Without diode pumping, solid state lasers tend to be inefficient and bulkier. Thus, diode-pumped solid state lasers represent a symbiosis of capabilities that has created fantastic opportunities which neither group of technologies could have enabled on their own. QW lasers are available over a wide wavelength range, with increased brightness, more user friendly packaging, improved reliability and most importantly, at lower costs [4-6]. The generation of green, blue and UV radiation by frequency upconversion in solid state lasers is also of great importance for various applications of high power QW lasers e.g. pumping of Ti-Sapphire lasers, replacement of copper vapour laser for pumping Dye lasers and many other applications [3].

Interestingly, the first laser developed in India was also a semiconductor laser! Though not very well known, in 1968 GaAs injection lasers were developed by a group at Bhabha Atomic Research Centre (BARC), Mumbai and were used to demonstrate day-time optical communication between BARC and Tata Institute of Fundamental Research (TIFR), Colaba, Mumbai [7,8]. In spite of these early achievements, unfortunately semiconductor laser development within India did not grow rapidly. Later, research groups at TIFR, Mumbai and SSPL, New Delhi fabricated QW lasers at several wavelengths [9-11]. Recently, The Semiconductor Laser Section at RRCAT, Indore has developed high power QW lasers in the wavelength range of 740 to 1000 nm. The complete laser structure is grown by metal organic vapour phase epitaxy (MOVPE) technique and devices are fabricated through standard procedure using photolithography. In this article, we summarize the progress of high power QW laser developmental work at RRCAT.

## 2. Experimental details

The laser structures were grown in a horizontal MOVPE reactor (AIX-200) with a rotating substrate holder on $n^+$ GaAs epi-ready substrates. Trimethyl gallium (TMGa), Trimethyl aluminum (TMAl), Trimethyl indium (TMIn), arsine ($AsH_3$) and Phosphine ($PH_3$) were used as the precursors. 2% Silane ($SiH_4$) in Hydrogen and Dimethyl Zinc (DMZn) were used as the dopant source for n and p type doping respectively. The typical growth rate for different layers in the laser structures varies from 2 to 6Å/sec at a reactor pressure of 50mbar. The total flow in the reactor was kept about 8 slpm. It is to be noted that we used epi-ready $n^+$ GaAs wafers only and hence did not find a necessity of any pre-cleaning or etching steps before loading GaAs substrates into the MOVPE reactor. A stereo zoom microscope was used for a quick imaging of surface morphology of the grown laser structures. Furthermore, better images at higher magnification were recorded using a scanning electron microscope (SEM).

GaAsP strained QW with AlGaAs barriers were also utilized to develop QW lasers operating at 740 and 810 nm. For the development of QW lasers operating at about 965 nm, InGaAs QWs with GaAs barriers were used. The compositional changes were also made in the respective waveguide and cladding layers for the particular lasing wavelength. High resolution X-ray diffraction (HRXRD) measurements were performed in order to measure the composition and thickness of epitaxial layers. HRXRD measurements are performed by using an X'PERT Pro MRD diffractometer, equipped with a Ge (220) monochromator (Bartels type, 4 reflections), with beam divergence of ~12 arcsecond in the scattering plane for CuK$_{\alpha 1}$ X-rays ($\lambda$=1.5406 Å). All HRXRD measurements are done with a 1° open detector in a plane parallel to the scattering plane [12]. Doping studies were performed for individual layers by Hall and electrochemical capacitance voltage (ECV) measurements. The complete laser structures were profiled for doping studies by ECV technique. ECV profiling is carried out using different electrolytes like Tiron, NaOH:EDTA and 1M $H_2SO_4$ in PN4300 setup. Initially, the respective QW structures were optimized through Photoluminescence (PL) and Surface photovoltage spectroscopy (SPS) measurements. PL was excited with a suitable laser beam ($\lambda$=325, 532, 660 or 810 nm depending on the bandgap of the respective material), dispersed with a 1/4 m monochromator and detected by either a photo multiplier tube or photodiode (Silicon or Germanium) using conventional lock-in technique [13]. The band pass of the monochromator is kept about 1nm. SPS measurements were performed in a chopped light geometry under soft contact mode [14,15]. After the MOVPE growth of laser structures a quick turn around processes based on the shadow mask technique is used for initial testing. Furthermore, the devices are fabricated through standard procedure using photolithography and lift-off technique where better stripes with sharp edges could be achieved.

## 3. Results and discussion

In this section, we present our results related to the optimization of QW structures through various characterization techniques. Device processing of high power QW laser diodes and there characteristics are also presented.

### 3.a MOVPE growth optimization of QW lasers

The very first optimization during the development of high power QW laser diodes is the fine tuning of wavelength of QW luminescence. Figure 1 shows the results of such an exercise utilized for the development of GaAs/AlGaAs based QW laser diodes operating at about 810 nm. Quantum confinement effects are clearly seen when the QW thickness is varied from 30 to 190 Å. We understand from this exercise that the thickness of GaAs QW need to be kept about 45 Å for the development of 810 nm laser diodes. Following this a complete laser structure with AlGaAs barrier and cladding layers was grown. A schematic layer diagram of 810nm laser diode structure is shown in Fig. 2. It has a double QW configuration working as the active region of the laser diode.

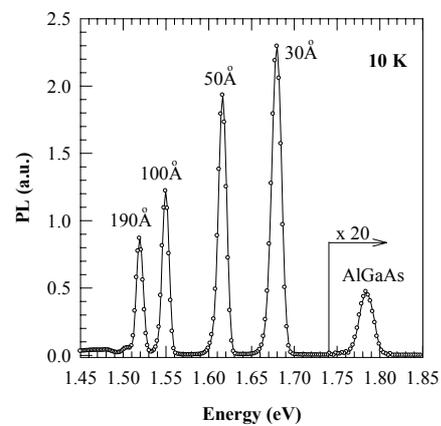

Fig.1 10K PL of a GaAs/AlGaAs multi QW sample with QW thickness varying from 30 to 190 Å.

| 400 nm p$^+$ GaAs (cap) |
| --- |
| 1.5 µm p Al$_{0.55}$Ga$_{0.45}$As(clad) |
| 110 nm Al$_{0.3}$Ga$_{0.7}$As(waveguide) |
| 4.5 nm GaAs QW |
| 15 nm Al$_{0.3}$Ga$_{0.7}$As (barrier) |
| 4.5 nm GaAs QW |
| 110 nm Al$_{0.3}$Ga$_{0.7}$As(waveguide) |
| 1.5 µm n Al$_{0.55}$Ga$_{0.45}$As(clad) |
| 0.2 µm n$^+$ GaAs |
| n$^+$ GaAs (substrate) |

Fig.2 The Schematic layer structures of a QW laser operating at about 810 nm

Figure 3 shows the comparison of PL and SPS spectrum of 810 nm laser structure shown in Fig.2. Curve 1 shows the room temperature PL spectrum of as grown laser structure. We observed a broad PL band originating from cap p$^+$ GaAs layer and no signature of GaAs QW was observed. After this, the GaAs cap layer was removed by chemical etching. Curve 2 shows the room temperature PL spectrum of laser structure after removal of cap GaAs. We observed two intense and sharp features which are related to 4.5 nm thick GaAs QW. Curve 3 shows the SPV spectrum of the same laser structure without any etching treatment. Several interesting features are observed as clearly seen from curve 3. The sharp QW features seen in PL spectrum are also reproduced by SPS technique even for as grown laser structure. In addition to this, there are two more features seen in SPV spectrum. The feature at about 692 nm is related to AlGaAs barrier layer. There is one more feature seen at about 739 nm which is again related to higher level of GaAs QW. We identify various transitions seen in PL and SPV spectrum by numerically solving the Schrödinger equation for a square potential well in effective mass approximation. We label three transitions of QW to $e_1$-$hh_1$, $e_1$-$lh_1$ and $e_2$-$hh_2$. Hence, it is clear that SPS provides additional information about the energy position of higher level of QW and AlGaAs barrier layer which we couldn't get in PL spectrum even after etching the cap GaAs layer. It shows the strength of SPS technique over conventional PL spectroscopy.

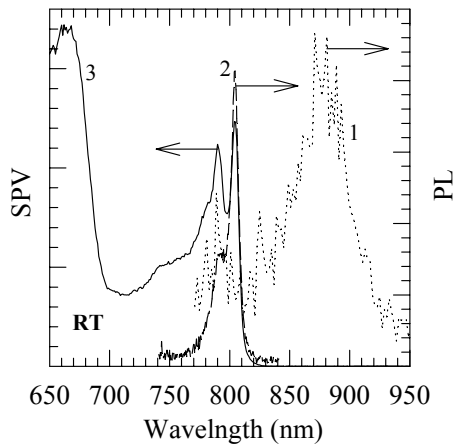

Fig.3. PL and SPV spectrum of as grown laser structure (curve 1 & 3), PL spectrum of cap layer etched laser structure (curve 2)

For the development of QW lasers operating at about 965 nm, $In_xGa_{1-x}As$ QWs with GaAs barriers were used. Figure 4 shows the 300K PL spectrum of a representative InGaAs QW sample emitting at 995nm. For the fine tuning of wavelength of QW luminescence, it is necessary to measure the thickness and composition of InGaAs QW. It is well established that HRXRD and PL correlation acts as an accurate and nondestructive evaluation technique for InGaAs/GaAs QW structures. Figure 5 shows ω/2θ diffraction patterns for the (004) reflection of InGaAs/GaAs QW samples with different indium content $x_s$. The scans are vertically shifted in figure for clarity in viewing. The intense sharp peak at the center is from GaAs substrate and the broad peak at low angle side is from InGaAs QW. The substrate peak is narrow and has the highest intensity while the QW peak is very broad and has very low intensity. As the indium content of the QW is increased, the QW feature moves towards lower ω/2θ values. The superposition of x-rays reflected from the interfaces gives fringes in the diffraction patterns. The separations between fringes are very sensitive to cap layer thickness. For each QW numerical simulations were performed and the values of $x_v$, $x_s$ and $L_z$ are shown in Table 1. We have also numerically solved the Schrodinger equation for InGaAs/GaAs QW structures using envelope function approximation. Effect of strain in modifying the band structure has also been taken into account. Indium content and thickness of QW samples are obtained from HRXRD measurements. The measured and calculated values of the lowest electronic feature of QW structures are shown in Table 1. From a close matching of the experimental and simulated values, it is obvious that the HRXRD results are strongly supported by PL measurements. As obvious from Table 1 that the indium content of InGaAs/GaAs QW increases with increase of trimethyl indium flux in vapour phase and the values of $x_s$ versus $x_v$ are plotted in Figure 6. The values of $X_v$ need to be modified by using an ultrasonic cell which measures the actual partial pressure of TMIn. These values are shown in brackets in Table.1 for three samples and are lower than the values simply calculated using flows and vapour pressures. It is possible to choose an appropriate value of trimethyl indium flux in the vapour phase in order to obtain a particular composition of InGaAs/GaAs QW laser structures which may operate at a desired wavelength.

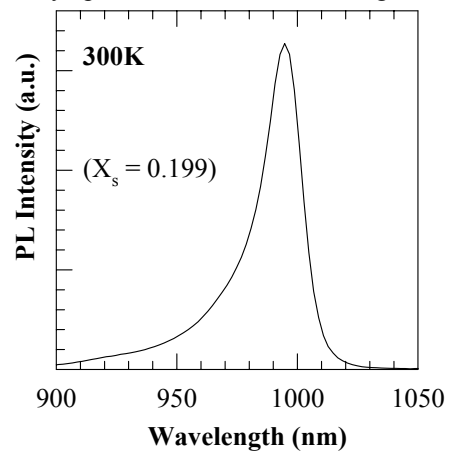

Fig. 4 300K PL of a InGaAs/GaAs single QW

### 3.b Device processing of QW lasers

Semiconductor laser diode processing is very involved and cumbersome process. For quick testing of laser diodes and checking its suitability at a particular wavelength, a quick turn around process based on the shadow mask technique can be used. To accomplish this we have made metal mask consisting 10-15 slits of about 150μm width with 500micron spacing in rectangular geometry on a very thin (~200μm) aluminum alloy sheet. The fabrication of μm size slit in this plate is a difficult task and this has been accomplished with laser metal cutting under optimized conditions. High power QW lasing was demonstrated using this scheme at several wavelengths in order to obtain a quick check on the laser structure quality. However, the shadow mask technique has an inherent limitation due to the nonuniform stripe edged defined by the laser metal cutting. Better stripes with sharp edges could be obtained by photolithography and lift-off technique. Therefore, we used photolithography and lift-off as a standard procedure for defining lateral dimension of the laser diodes. The front contacts to the p+ GaAs cap layer of the laser structures were made by depositing Ti/Pt/Au (200Å/1000Å/3000Å) using a metal coating unit at $3\times10^{-6}$ mbar pressure. The metal coated samples were then boiled in acetone briefly to lift-off the unwanted metal with underlying resist. To isolate each device, the laser structures were etched till the cladding AlGaAs layer using $H_3PO_4:H_2O_2:H_2O(1:1:8)$ solution. To reduce the series resistance of the device, n+ GaAs wafer was then thinned down to 125 μm by lapping and polishing. Au-Ge/Ni/Au (200Å/100Å/3000Å) was then deposited to make the Ohmic contacts to the back side (n+ GaAs). The laser diode was then annealed at 480 $^0$C to reduce the contact resistances to its minimum value. The typical value of the series resistance for these devices varies from 1-1.5 Ω. The whole wafer was then cleaved into 500 μm to 2 mm long laser bars using a manual scriber.

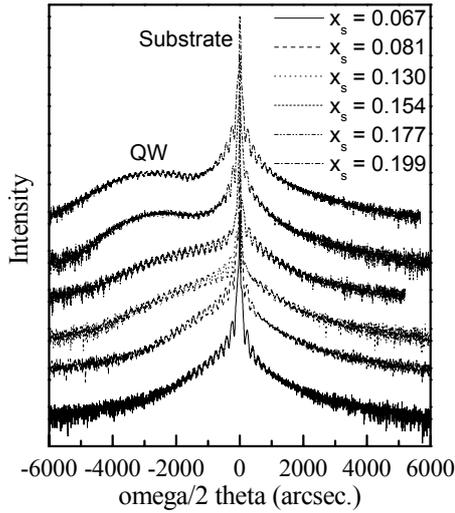
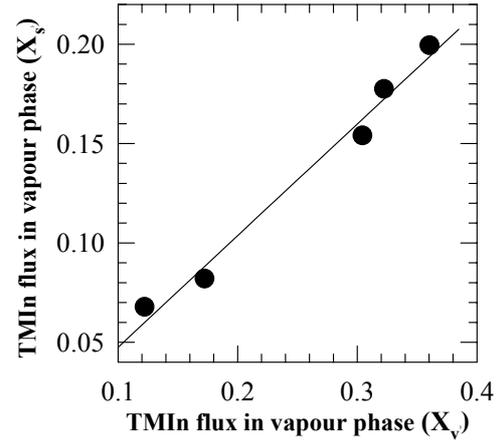

Fig. 5 HRXRD patterns for (004) reflection of InGaAs/GaAs QW samples.

Fig. 6 $X_s$ versus $X_v$ for InGaAs/GaAs QW samples.

| Sample No. | $x_v$ | $x_s$ | $L_z$ (Å) | PL peak positions (nm) at 300K | |
|---|---|---|---|---|---|
| | | | | Experimental | Simulation |
| 1 | 0.123(0.094) | 0.067 | 78 | Not measured | 906 |
| 2 | 0.173(0.128) | 0.081 | 80 | 910 | 908 |
| 3 | 0.305(0.218) | 0.154 | 71 | 958 | 958 |
| 4 | 0.323(NM) | 0.177 | 82 | 974 | 973 |
| 5 | 0.361(NM) | 0.199 | 76 | 995 | 995 |

Table 1. Values of $x_v$, $x_s$, $L_z$ obtained from HRXRD measurements and experimental and simulated PL peak positions for InGaAs/GaAs QW samples. The actual partial pressure values of TMIn using an ultrasonic cell are also shown in brackets for three samples.

### 3.c Device testing of QW lasers

The laser bars were tested using a homemade laser diode driver and an OPHIR power meter. A typical light output vs. driving current (L-I) characteristics of a 100 μm wide and 1 mm long laser diode emitting at 851 nm wavelength are shown in Fig. 7. L-I measurements were performed under uncoated, unmounted conditions. The peak power obtained in pulsed operation is about 5.3 W which is further limited by the capacity of laser diode driver. These laser diode structures were tested under pulsed operation at room temperature for 500 nanosecond pulse duration with a duty cycle of 1:1000. The devices died at a peak power of about 4.2W due to heating when tested with 1 μs pulse duration with a duty cycle of 1:500. Lasing action was observed with a typical threshold current density of about 200A/cm$^2$. Laser diode arrays consisting of 6-10 elements were also developed and tested for operation in pulsed mode at room temperature. Figure 8 shows a laser array with 6 elements under operation very close to the current threshold at room temperature. These were installed in p-side down geometry where a metal single tip-electrode was used for back electrical connection during operation. Figure 9 shows the longitudinal spectrum of six laser diodes operating at different wavelengths.

## 4. Conclusion

In summary, we have demonstrated high power QW lasers indigenously developed in our lab in the wavelength range of 740 to 1000 nm. Several issues related to epitaxial growth, device processing and testing have been addressed. However, there are a few more essential device processing steps like facet coating, bonding and packaging of laser diodes which need to be implemented before we could operate our device under cw operation. Successful implementation of these steps will ensure an efficient heat removal from the chip and the usefulness of indigenously developed devices. Efforts are continuing in this direction.

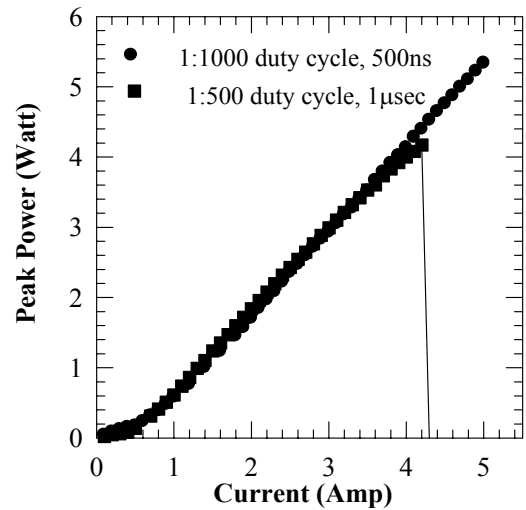

Fig.7 L-I characteristics of a laser diode

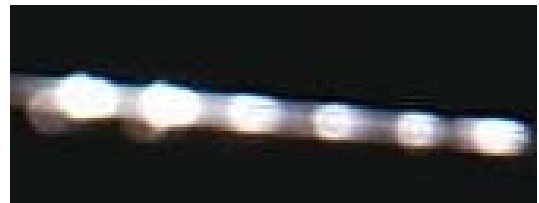

Fig.8 A Laser diode array with 6 elements

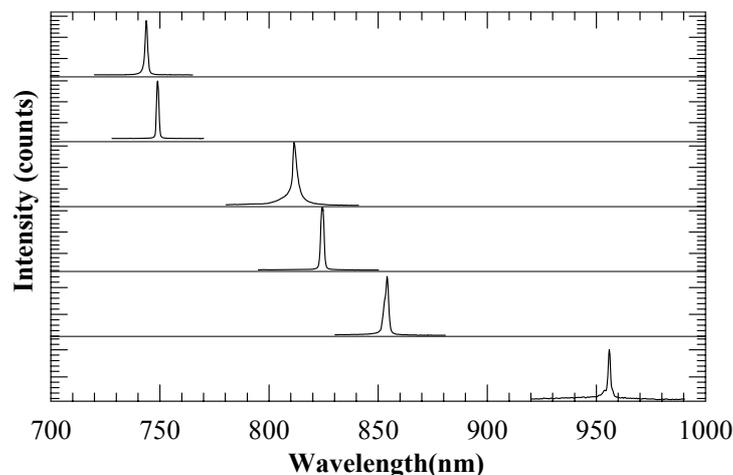

Fig. 9 Longitudinal spectra of six laser diodes operating at different wavelengths

**5. Acknowledgments:** Authors would like to thanks Dr. V. C. Sahni, Prof. B. M. Arora and Prof. K. C. Rustagi for the encouragement and support and Shri U. K. Ghosh for the technical support.

## 6. References

1. N. Holonayak, Jr., IEEE J. Selected Topics Quant. Electron., **6**, (2000)
2. P. Zory (Ed), Quantum Well Lasers, Academic Press, Boston, 1993.
3. R. Diehl (Ed), High Power Diode Lasers, Springer, Berlin, 2000.
4. Markus Weyers, "GaAs-based high power laser diodes", 11th European Workshop on MOVPE, Lausanne June 5th-8th 2005.
5. T. K. Sharma and A. Bhattacharya, "Semiconductor Lasers- A Review Article", KIRAN, **13**(1), 7(2002).
6. T. K. Sharma et al., IEEE Photon. Technol. Lett. **14**(7), 887 (2002).
7. M. B. Khambaty et al., Ind. J. of Pure and Appl. Phys. **7**, 29 (1969).
8. U. K. Chatterjee et al., Ind. J. of Pure and Appl. Phys. **8**, 488 (1970).

9. T. K. Sharma, A.P Shah, M. R. Gokhale, C. J. Panchal and B. M. Arora, Semicond. Sci. Technol. Vol. **14**, 327 (1999).
10. S. S. Chandvankar, A. P. Shah, A. Bhattacharya, K. S. Chandrasekaran and B. M. Arora, J. Crystal Growth, **260**, 348 (2004).
11. S.K.Mehta et.al. Proc SPIE, 2733, 122, 1996.
12. P. F. Fewster and N. L. Andrew, J. Phys. D **28** (1995) A97.
13. T.K. Sharma, B. M. Arora, S. Kumar and M. R. Gokhale, J. Appl. Phys., **91** (9) 5875 (2002).
14. T. K. Sharma, S. Porwal, R. Kumar and S. Kumar, Rev. Sci. Inst. **73** (2002) 1835.
15. S. Datta, S. Ghosh, and B. M. Arora, Rev. Sci. Instrum. **72** (2001) 177.